\documentclass{article}
\usepackage{graphicx}
\def\espacefig{\bigskip}
\def\la{\langle}
\def\ra{\rangle}
\def\real{{{\rm I}\!{\rm R}}}
\def\1{{\rm Id}}
\def\lint{{\scriptstyle{\rfloor}}}

\newcommand\maath{\mathsurround=0pt}
\newcommand{\EQM}[1]{\vcenter{\normalbaselines\maath
    \ialign{${\displaystyle ##}$\hfil&&\ ${\displaystyle ##}$\hfil\crcr
    \mathstrut\crcr\noalign{\kern-\baselineskip}
    \noalign{\smallskip}
    #1\crcr\mathstrut\crcr\noalign{\kern-\baselineskip}}}}

\begin{document}

\centerline{\bf Projective dynamics and classical gravitation}
\centerline {Alain Albouy, albouy@imcce.fr}
\centerline {ASD/IMCCE-CNRS-UMR 8028}
\centerline{Observatoire de Paris}
\centerline {77, avenue Denfert-Rochereau, 75014 Paris}
\centerline{3/5/2005}
\bigskip\bigskip

{\sl Abstract.} Given a real vector space $V$ of finite dimension, together with a particular homogeneous field of
bivectors that we call a {\sl field of projective forces}, we define a law of dynamics such that the position of the
particle is a {\sl ray} i.e.\ a half-line drawn from the origin of $V$. The impulsion is a bivector whose support is a
2-plane containing the ray. Throwing the particle with a given initial impulsion defines a projective trajectory. It
is a curve in the space of rays ${\cal S}(V)$, together with an impulsion attached to each ray. In the simplest example
where the force is identically  zero, the curve is a straight line and the impulsion a constant bivector in
$\bigwedge^2 V$. A striking feature of projective dynamics appears: the trajectories are not parameterized.

Among the  projective force fields corresponding to a central force, the one defining the Kepler problem is simpler than
those corresponding to other homogeneities. Here the thrown ray describes a quadratic cone whose
section by a hyperplane corresponds to a Keplerian conic. An original point of view on the hidden symmetries of the
Kepler problem emerges, and clarifies some remarks due to Halphen and Appell. We also get the unexpected conclusion that
there exists a notion of divergence-free field of projective forces if and only if
$\dim V=4$. No metric is involved in the axioms of projective dynamics.

\bigskip

\centerline {\bf 1. Introduction}

The main object of the present study is the class of differential systems defining, on an open set $U$ of an affine space
$A$, the motion of a particle subjected to a field of forces. We write such a
system:
$$\ddot q=f(q),\qquad q\in U\subset A,\qquad f:U\to\vec A.\eqno(1.1)$$
The force or
acceleration $f(q)$ lives in the vector space $\vec A$ associated to $A$. An elementary computation,
apparently due to Appell, indicates that it makes sense to ``projectivize'' such systems, considering the
$n$-dimensional affine space $A$ as an affine hyperplane of a $n+1$-dimensional vector space $V$. Then
$A$ plays the role of an affine chart for
${\cal P}(V)$, the real projective space associated to $V$, or better for ${\cal S}(V)$, the double covering of ${\cal
P}(V)$, whose points are the rays from the origin $O$ of $V$. The topological aspects of this projectivization
are interesting, but quite trivial: one can sometimes extend the domain $U$ of the motion; for example one
``closes'' the hyperbolic orbits of the Kepler problem, making them ellipses. The local aspects are more surprising:
Appell's computation indicates that something as
$(1.1)$ is already defined when the particle lives in a space which is not affine and whose tangent bundle is not
endowed with a linear connection.

1.1. {\sl Appell's computation.} It may be presented as follows. As $A\subset V$, the ``point'' $q\in A$ is now a
``vector''. Let
$h\in V^*$ be the linear form such that $\la h,q\ra=1$ is the equation of $A$. We choose a non-zero $h_1\in V^*$ and
call
$A_1$ the affine hyperplane
of equation $\la h_1,q\ra=1$. To a
$q\in A$ with $\la h_1,q\ra>0$ we associate
$q_1=\la h_1,q\ra^{-1}q\in A_1$ on the same ray. We compute
$$\dot q_1=\frac{\la h_1,q\ra\dot q-\la h_1,\dot q\ra q}{\la h_1,q\ra^2}.$$
Because of the denominator, the second derivative $\ddot q_1$ is quite complicated. But we change the time
parameter. We define the derivative $r'$  of a quantity $r$ with respect to the new time by
$$r'=\la h_1,q\ra^2\dot r.\eqno(1.2)$$
Then
$$q_1''=\la h_1,q\ra^2\bigl(\la h_1,q\ra\ddot q-\la h_1,\ddot q\ra q\bigr).\eqno(1.3)$$
Substituting $\ddot q=f(q)$ and $q=\la h,q_1\ra^{-1} q_1$, we observe that the right hand side, which is  a vector tangent
to
$A_1$ at
$q_1$, only depends on the position 
$q_1$. Thus {\sl a system such as $(1.1)$, i.e.\ defined by a field of forces depending only of the position,
remains of the same type after a ``change of projection'', provided the time parameterization is changed according to the
rule $(1.2)$.}

\espacefig
\centerline{\includegraphics [width=50mm] {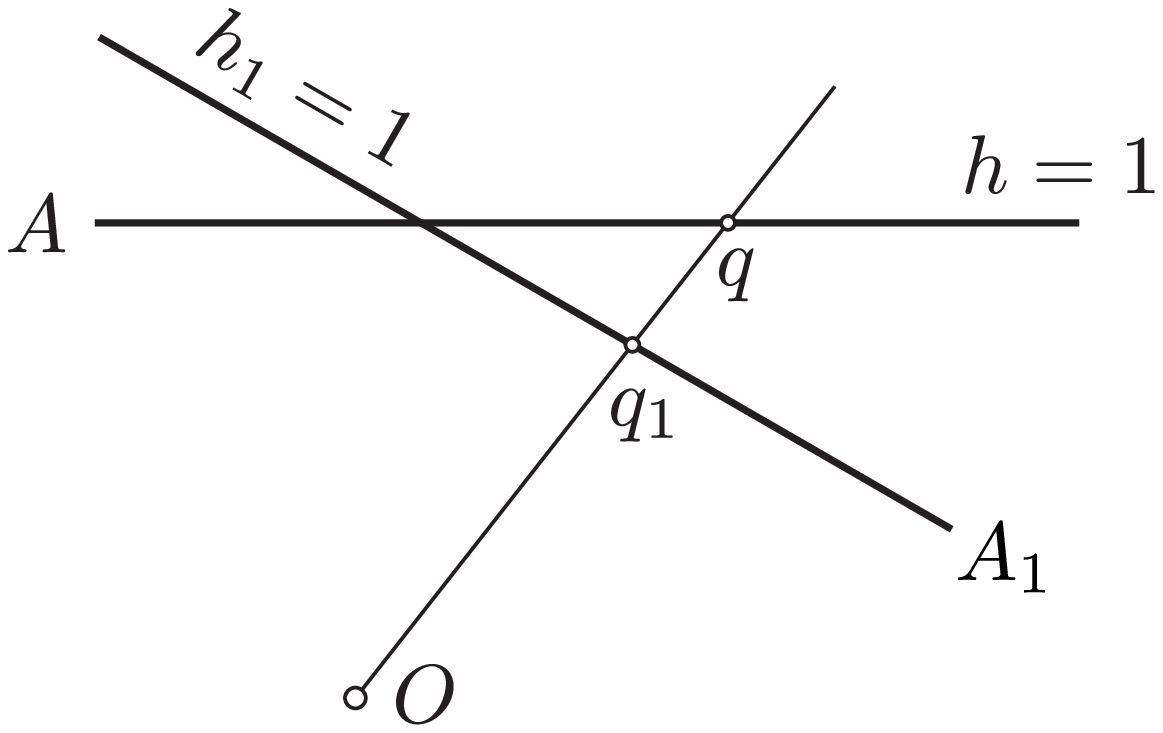}}
\nobreak
1.2. {\sl Figure}

1.3. {\sl Halphen's computation.} It particularizes the previous computation and introduces the
``classical gravitation'' aspect. We consider a center of force $c\in A$. We consider the particular case of
$(1.1)$
$$\ddot q=-\|q-c\|^\beta (q-c),\qquad q\in A\setminus\{c\},\qquad \beta\in\real.\eqno(1.4)$$
We endowed the affine space $A$ with a Euclidean structure $\|.\|^2$, making it a $n$-dimensional Euclidean space. We
apply the above transformation
$$\EQM{q_1''&=-\|q-c\|^\beta\la h_1,q\ra^2\bigl(\la h_1,q\ra(q-c)-\la h_1,q-c\ra q\bigr)\cr
&=-\|q-c\|^\beta\la
h_1,q\ra^2\bigl(-\la h_1,q\ra c+\la h_1,c\ra q\bigr)}$$
or$$q_1''=-\|q-c\|^\beta\la h_1,q\ra^3\la h_1,c\ra\Bigl(\frac{q}{\la
h_1,q\ra} -\frac{c}{\la h_1,c\ra}\Bigr).$$
We shall express the right hand side in term of $q_1=\la h_1,q\ra^{-1}q\in A_1$ and $c_1=\la h_1,c\ra^{-1}c\in A_1$.
For this we extend the Euclidean quadratic form $\|.\|^2$ from $\vec A$ to $V$ in a special way. We decompose any $q\in V$ in horizontal plus vertical components: $q=q_A+\lambda c$, where $q_A\in \vec A$ and
$\lambda\in\real$. Here we think of $\vec
A$ as the horizontal vector hyperspace in $V$, of equation $\la h,q\ra=0$.
We set $\|q\|^2=\|q_A\|^2$; in particular $\|c\|^2=0$, and $\|q-c\|^2=\|q\|^2$.  We resume the computation:
$$q_1''=-\|q\|^\beta\la h_1,q\ra^3\la h_1,c\ra (q_1-c_1)=-\|q_1\|^\beta\la h_1,q\ra^{3+\beta}\la h_1,c\ra(q_1-c_1).$$
As $\la h,q_1\ra=\la h_1,q\ra^{-1}$, we get
$$q_1''=-\|q_1-c_1\|^\beta\la h,q_1\ra^{-3-\beta}\la h_1,c\ra(q_1-c_1).\eqno(1.5)$$
{\sl The transformed system $(1.5)$ has the
same form as the original system $(1.4)$ if  $\beta=-3$, i.e.\ if $(1.4)$ defines the Kepler problem and corresponds
to the Newtonian attraction.} Actually we must slightly extend the form $(1.4)$ to accept the ``mass'' factor $\la
h_1,c\ra$ in $(1.5)$. Moreover, the affine hyperplane $A_1$ is endowed with the restriction of the degenerate
quadratic form
$\|.\|^2$ to it. If $A_1$ is ``vertical'', i.e.\ if it contains the vertical vector $c$, this restriction is
degenerate; again we shall slightly extend the class of equations $(1.4)$ to accept this kind of $\|.\|^2$.

It is important to have clearly in mind that if $\beta\neq -3$  the transformed system $(1.5)$ is more complicated
than $(1.4)$.

1.4. {\sl Divergence of the field of forces.} The next statement could be obtained by a long but
straightforward computation. We will prove it simply in \S 5. {\sl If
$\dim A=3$ and
${\rm div} f=0$ in
$(1.1)$, then the field of forces in the right hand side of $(1.3)$, expressed in the new variable $q_1$, is also
divergence-free.}

This statement is readily checked in the example of the Kepler problem. It is well-known that in dimension 3 the force
function in this problem satisfies the Laplace equation, and that consequently its gradient, the field of forces,
is divergence-free.

The Kepler problem has another striking property: its bounded orbits are periodic. Newton emphasized this property.
For him one tests the inverse square law verifying that the aphelia of Saturn and the comets are at rest. This
law is $(1.4)$ with $\beta=-3$; the force field is divergence-free. Is this coexistence between the dynamical fact
and the null divergence a mere coincidence? We do not know if this question has already been discussed. Maybe it seems
too vague, especially if we stick to the unique Kepler problem and do not discuss other examples. In the 19th century,
examples with both the dynamical and the divergence-free properties were discovered. They are discussed in [Alb],
[AlS] and [BoM]. We will present the most general example we know in \S 6.

Does projective dynamics explain this coexistence? Choosing the projective point of view we ``forget" the affine and
metric structures of the space. But we keep a structure which is sufficient to express the
divergence-free property, the central force property and the degeneracy of the dynamics. So, if some explanation is to be
found, it is reasonable to look for it inside projective dynamics. Actually when we think that there exists a relation
between these properties, we immediately raise two objections. The first one remains for us mysterious: the dynamics is
two-dimensional, while the divergence-free hypothesis is three-dimensional.

A second objection is rather a question: does a field of forces that is divergence-free in dimension $n$
define a special dynamics? Or is there something special only if $n=3$? A deep property of the case
$n=4$ was discovered by Jacobi (see [AlC] pp.\ 161 and 169, [Mon]). Concerning the central force field which is
divergence-free in dimension $n=2$, we do not know anything remarkable, except that Johann Albrecht Euler, son of
Leonhard Euler, obtained an expression for the total duration of a rectilinear free fall, with a zero initial velocity. As
mentioned to me by James E.\ Howard, this result is an elementary exercise requiring the computation of
$\int_0^\infty e^{-v^2/2}dv$.

Projective dynamics offers a surprising answer to this second objection: in this
``poorly structured'' context, the concept of a divergence-free field of forces only exists in dimension $3$.

\bigskip

\centerline{\bf 2. $s$-tangent vectors and $s$-scalars}

As above,  $V$ is a real vector space of
dimension
$n+1$. A
{\sl ray} is an open half-line drawn from the origin in $V$. In projective dynamics, the position of a particle is a
ray.

We
call ${\cal S}(V)$ the manifold of the rays of $V$.
To describe in words and symbols a relation between $V$ and ${\cal S}(V)$, we use such expressions as:
a point $Q\in{\cal S}(V)$ {\sl corresponds} to a ray $\hat Q\subset V$. Actually there are two presentations of the
projective facts. One is $n+1$-dimensional and elementary; it corresponds to the classical homogeneous coordinates.
The other one is $n$-dimensional, which is more natural; but it leads to abstract constructions.

2.1. {\sl Definition.} Let $s\in\real$. A {\sl $s$-tangent vector} $w$ to ${\cal S}(V)$ at $Q\in{\cal S}(V)$
corresponds to a class of homogeneous vector fields of degree $s$ along the ray $\hat Q\subset V$. We define a class
$\hat w$ as follows. Let $w_0:\hat Q\to V$ be a homogeneous vector field in $\hat w$. Then $w_1:\hat Q\to V$ is in the
same class if and only if the vector field
$w_1-w_0$ is tangent to the ray $\hat Q$.

Given an open set ${\cal U}\subset {\cal S}(V)$ we denote by $T^s_Q{\cal U}$ the $n$-dimensional vector space of
$s$-tangent vectors at $Q\in{\cal U}$. The description of a tangent vector to ${\cal P}(V)$ or to ${\cal S}(V)$ as a
$1$-tangent vector is standard. It is the identity $T^1_Q{\cal U}=T_Q{\cal U}$. For the values
$s\neq 1$ of the homogeneity, a non-zero $s$-tangent vector is not a tangent vector, but nevertheless it points a
direction tangent to ${\cal U}$.

2.2. {\sl Equivalent definition.} Let $s\in\real$. A {\sl $s$-tangent vector} $w$ to ${\cal S}(V)$ at $Q\in{\cal S}(V)$
corresponds to a  $\hat w: \hat Q\to \bigwedge^2 V$ satisfying $q\wedge
\hat\omega (q)=0$ for any $q\in\hat Q$, and
$\hat w(\lambda q)=\lambda^{s+1}\hat w(q)$ for any $\lambda>0$.

We defined $\hat w$ first as a class of homogeneous vector fields along the
ray
$\hat Q\subset V$, then as a homogeneous bivector field along $\hat Q$. It is easy to relate both definitions. If
$w_0:\hat Q\to V$ is in the class $\hat w$, we associate to it the bivector field $q\wedge w_0(q)$, which is
homogeneous of degree
$s+1$ and satisfies $q\wedge q\wedge w_0(q)=0$. If we take another vector field $w_1$ in the same class, then $q\wedge
w_0=q\wedge w_1$. Conversely we consider
$\hat w:\hat Q\to\bigwedge^2 V$ and we choose any
$q\in\hat Q$. As $q\wedge\hat\omega(q)=0$ there exists a $v\in V$ such that $\hat \omega(q)=q\wedge v$.
We set $w_0(q)=v$ and extend $w_0$ by homogeneity.

The notion of a $s$-tangent vector is extremely useful in projective dynamics. This is why we introduced it first.
Nevertheless the following definition of a generalized ``scalar'' quantity should logically come first.

2.3. {\sl Definition.} Let $s\in\real$. A {\sl $s$-scalar} $\rho$ at $Q\in {\cal S}(V)$ corresponds to a function
$\hat\rho :\hat Q\to\real$ such that $\hat \rho(\lambda q)=\lambda^s\hat\rho(q)$ for any $q\in\hat Q$ and
any $\lambda>0$.

A $0$-scalar  is simply a real number. For a ${\cal U}\subset {\cal S}(V)$ we denote by
$\theta^s{\cal U}$ the line bundle (i.e.\ the vector bundle with one-dimensional fiber) of $s$-scalars. The notation
${\cal O}(s)$ for the same object is widely used in the case of a complex projective space,  $s$ being an integer. The
following formulas are quite standard: 
$$\theta^s{\cal U}=(\theta^1{\cal U})^{\otimes s},\quad
\theta^r{\cal U}\otimes \theta^s{\cal U}=\theta^{r+s}{\cal U},\quad\theta^s{\cal U}\otimes T^r{\cal U}=T^{r+s}{\cal
U}.$$
Here the tensor products of fields defined on ${\cal S}(V)$ {\sl correspond} to mere products of the corresponding
fields defined on
$V$.

\bigskip
\centerline{\bf 3. The data used in projective dynamics}

3.1. {\sl Definition.} We  call a $(-1)$-tangent vector a {\sl projective impulsion} and a $(-3)$-tangent vector a {\sl
projective force}.

A field $F$ of projective forces on ${\cal U}\subset {\cal S}(V)$ is a section of the vector
bundle $T^{-3}{\cal U}$. With a standard notation this reads $F\in\Gamma(T^{-3}{\cal U})$. To
describe the ``corresponding'' object, we denote by
$\hat {\cal U}\in V$ the union of the rays corresponding to the points of ${\cal U}\subset{\cal S}(V)$. The field
$F$ corresponds to a map
$\hat F:\hat{\cal U}\to \bigwedge^2 V$, positively homogeneous of degree $-2$ and such that $q\wedge \hat F(q)=0$ at
any
$q\in
\hat {\cal U}$.

A {\sl parameterized path} is a (smooth or real analytic) map $\varphi: I\to {\cal U}$ where $I\subset\real$ is an open
interval. To define an {\sl oriented (unparameterized) path}, we simply weaken the structure of the source space $I$.
 
3.2. {\sl Definition.} An
{\sl oriented path} is a map $c: {\cal I}\to {\cal U}$, where ${\cal I}$ is a 1-dimensional oriented manifold
diffeomorphic (or analytically diffeomorphic) to $\real$.

While working with autonomous systems as $(1.1)$ it is natural to consider that a solution is more than an oriented
path, but less than a parameterized path. We mean we do not consider that the scalar value of the time $t$ at a
point of the path is a relevant information; but the time
$\Delta t$ to go from a point to another is well-defined. We call such solution a {\sl usual
trajectory}.

It is useful for our purpose to describe a usual trajectory in the following way: {\sl it is an oriented path together
with a field of velocity vectors}. The velocity vector at a point must be tangent to the path at this point. To take
into account the possibility of singularities and multiple points, we should endow the source manifold ${\cal I}$,
rather than the image of $c:{\cal I}\to {\cal U}$, with a field of non-zero, positively oriented vectors. Such data
induces a map
${\cal I}\to T{\cal U}$.

{\sl A projective trajectory is an oriented path together with a field of tangent projective impulsions.} This is the
main idea. However the possibility of multiple points and singularities somewhat complicates the precise definition.

\espacefig
\centerline{\includegraphics [width=100mm] {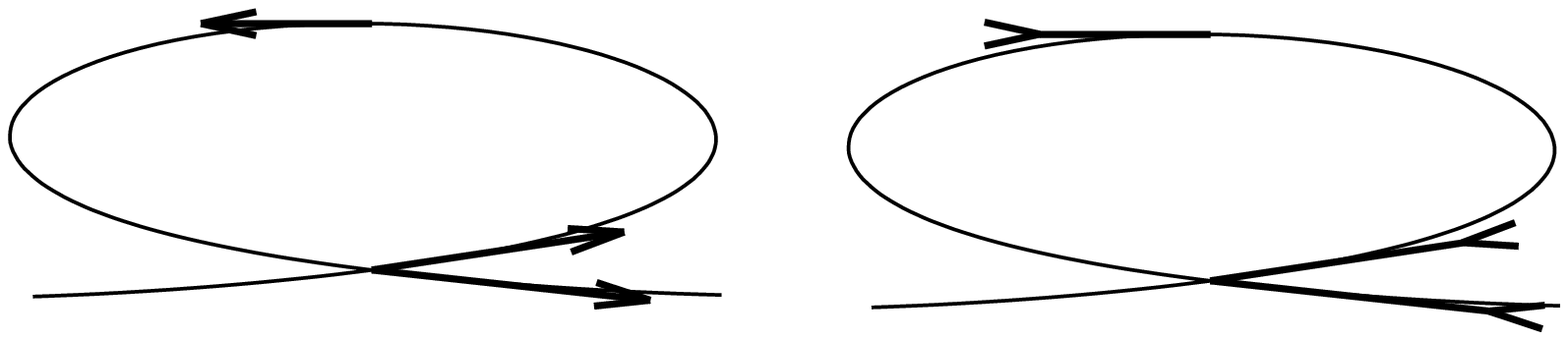}}
\nobreak
3.3. {\sl Figure.} A usual trajectory and a projective trajectory. An attempt of visualization.

3.4. {\sl Definition.} Let $c:{\cal I}\to {\cal U}$ be an oriented path. Let $\Theta{\cal I}$ be the line bundle
above ${\cal I}$, pull-back by $c$ of $\theta^{-2}{\cal U}$. A
structure of projective trajectory on
$c$ is a positive section of the oriented line bundle
$T{\cal I}\otimes\Theta{\cal I}$. A {\sl projective trajectory} is an oriented path together with a structure
of  projective trajectory.

3.5. The exponent $-2$ in the definition of the bundle $\Theta{\cal I}$ is the difference between the homogeneity of
a projective impulsion and the homogeneity of a tangent vector. To put this abstract definition into practice, let us
show that a projective trajectory
$(c,\sigma)$, where
$c:{\cal I}\to {\cal U}$ and
$\sigma\in\Gamma(T{\cal I}\otimes\Theta{\cal I})$, induces a map to the projective impulsions
$c_\sigma:{\cal I}\to T^{-1}{\cal U}$. At
$x\in{\cal I}$ we choose any non-zero $\rho\in T^*_x{\cal I}$, which determines a unique $v\in T_x{\cal I}$ such that
$\la \rho,v\ra=1$. The contracted product $\rho\cdot\sigma|_x\in\Theta_x{\cal I}$ defines  a
$(-2)$-scalar
$\lambda\in\theta^{-2}_{y}{\cal U}$, where $y=c(x)$. Let us call $c_*v\in T^1_{y}{\cal U}$ the push-forward of $v$.
The resulting projective impulsion  $c_\sigma(x)=\lambda\otimes c_*v\in T^{-1}_{y}{\cal U}$ does not depend on the
choice of $\rho$: if we multiply $\rho$ by a non-zero real number $\alpha$, we multiply $\lambda$ by $\alpha$ and
divide $v$ by $\alpha$.

\espacefig
\centerline{\includegraphics [width=95mm] {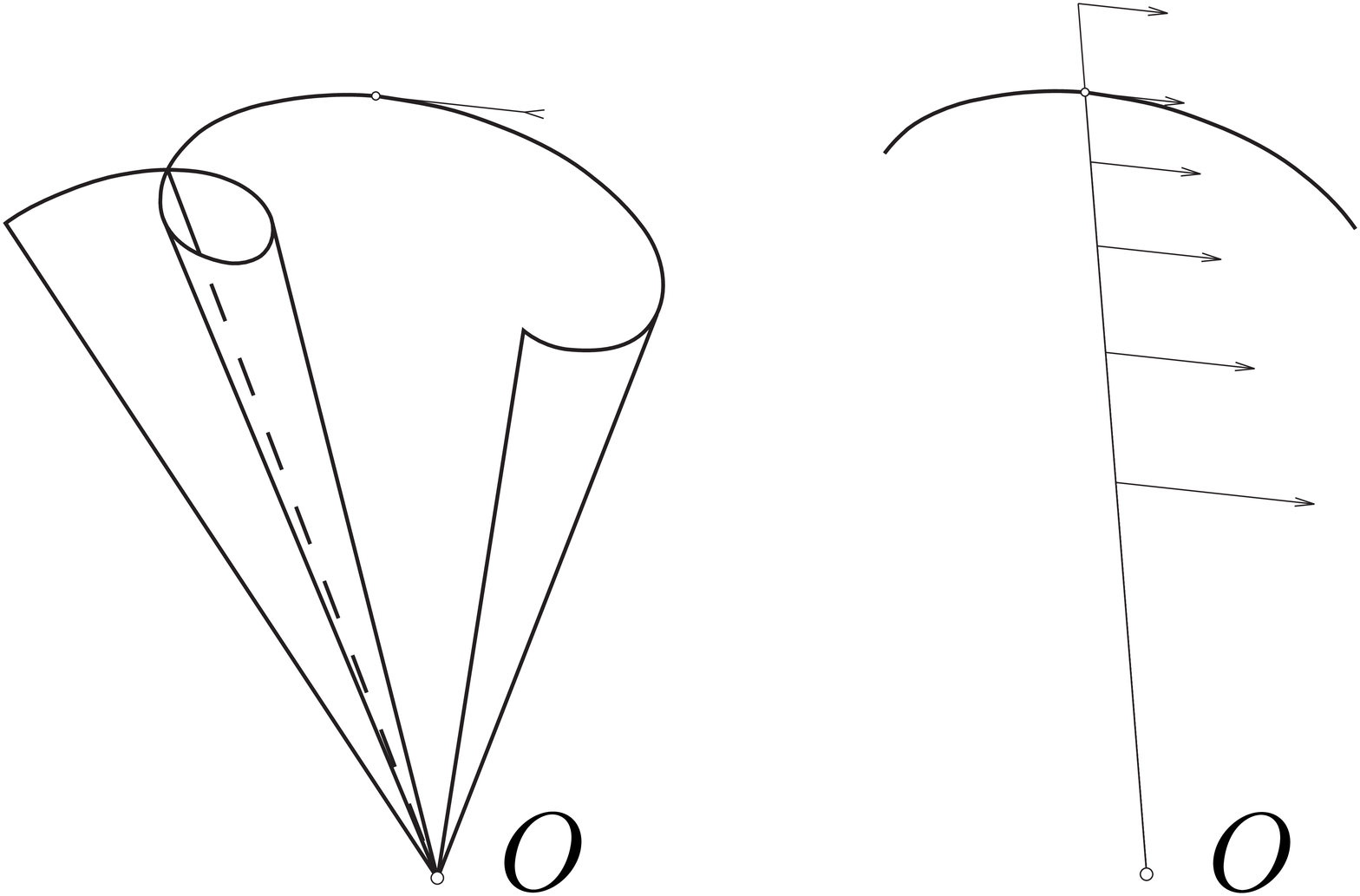}}
\nobreak
3.6. {\sl Figure.} A  projective trajectory and a projective impulsion.

As always the definition of the corresponding object is more elementary. To an oriented path $c:{\cal I}\to {\cal U}$
corresponds a two-dimensional ``half-ruled'' submanifold of ${\cal I}\times \hat{\cal U}$. By half-ruled we mean that
it contains half-lines instead of lines. The projection of this object on the factor $\hat{\cal U}$ is a
two-dimensional semi-cone, possibly with multiple points and singularities. A structure of projective
trajectory is a field of bivectors tangent to this semi-cone, possibly many-valued. The ``tangency condition'' must
take into account the possible singularities of $c$.

3.7. In the definition below and everywhere in this text,  ${\cal I}$ is an oriented manifold diffeomorphic to $\real$.
We denote by $I$ an open interval, and
$\tau: I\to {\cal I}$, $t\mapsto\tau(t)$ a global chart of ${\cal I}$, which respects the orientation.

3.8. {\sl Equivalent definition.} A {\sl projective trajectory} corresponds to an oriented path $c:{\cal I}\to {\cal
U}$ together with a map
$  \pi: {\cal I}\to \bigwedge^2V$ satisfying the following ``tangency condition'': there exist a global chart $\tau: I\to
{\cal I}$ and a map $q: I\to V$ satisfying $q(t)\in\hat Q(t)$, where $Q(t)=c(\tau (t))$, and 
$\pi(\tau(t))=q(t)\wedge\dot q(t)$.

3.9. To show that both definitions are equivalent, we start from a projective trajectory $(c,\sigma)$ in the sense of
Definition 3.4. We construct $c_\sigma$ as in 3.5 and set $\pi=\hat
c_\sigma$. We shall check that $\pi$ satisfies at any
$x\in {\cal I}$ the ``tangency condition''. Let $\tau: I\to {\cal I}$ be a chart 3.7. We set
$\tau(t)=x$,
$Q(t)=c(x)$, and construct with $\rho=dt$ a $\lambda\in\theta^{-2}_Q{\cal U}$ as we did in 3.5. Then ${c_*v}$
corresponds to a
$\hat{c_*v}:\hat Q\to\bigwedge^2 V$ such that $\hat{c_*v}(q(t))=q(t)\wedge\dot q(t)$ as soon as $q(t)\in\hat Q(t)$. We
select the unique
$q(t)$ in the ray $\hat Q(t)$ such that $\hat\lambda(q(t))=1$. The tensorial product relation in
3.5 means that $\pi(x)=\hat{c_*v}(q(t))$, which gives the result. The same computations allow to construct the section
$\sigma$ if we start with Definition 3.8 of a projective trajectory.

3.10. Why did we choose $s=-1$ for the projective impulsion? We will see that it is
forced by our introductory remarks. But we have already a strong argument to present. Among the projective trajectories
we can distinguish the linear ones, which ``draw'' part of a projective line in ${\cal S}(V)$. Among
these rectilinear motions we should be able to distinguish the ``uniform motions''. The statement ``constant
projective impulsion'' sounds good, and it makes sense with our choice $s=-1$. In this case, by the ``bivector
definition'' of a
$s$-tangent vector, the bivector is constant along the rays. To define the uniformity of the motion, we
require that the bivector is constant on the whole vectorial plane of $V$ that corresponds to the rectilinear
trajectory in ${\cal S}(V)$.

\bigskip

\centerline{\bf 4. From screen dynamics to projective dynamics}

In this section, the statement that occupies us is:
{\sl in a given field of projective forces, a ``position ray'' and a projective impulsion at this ray uniquely determine
a projective trajectory.} To establish it, we  study the process that converts our quite abstract ``projective
dynamics'' into a more familiar ``screen dynamics'', whose simplest case is $(1.1)$. The converted statement is the
usual statement that an initial position together with a velocity uniquely determine a trajectory. We convert again this
trajectory into a projective trajectory. Finally we prove that the resulting projective trajectory is independent of the
screen used in its construction (except if the screen is too small, i.e.\ if it covers only part of the
projective trajectory).

4.1. {\sl Definition.} Let ${\cal U}\subset {\cal S}(V)$ be an open set and $\hat {\cal U}\subset V$ the corresponding
open semi-cone. A {\sl screen function} $h$ for
${\cal U}$ is a positive $h:\hat {\cal U}\to \real$, positively homogeneous of degree one\footnote{Thus it
corresponds to a positive section of the $1$-scalar bundle $\theta^1{\cal U}$. There is no special reason to choose one
as the degree of homogeneity. We simply need a convention. }. The associated {\sl screen}
${\cal U}_h$ for
${\cal U}$ is the hypersurface of $\hat {\cal U}$ with equation $h=1$.

The simplest screen function is a linear form on $V$. It gives a ``flat'' screen for an open ``hemispheric'' ${\cal
U}\subset{\cal S}(V)$. In \S1 we used a pair $(A,A_1)$ of flat screens.

4.2. {\sl Restriction of $s$-tangent vectors to a screen.} Consider the screen $h=1$ for ${\cal U}\subset {\cal S}(V)$.
A $s$-tangent vector
$v$ at $Q\in{\cal U}$ defines a unique vector $v_h(q)\in V$, tangent to the screen at the unique $q\in \hat Q$  with
$h(q)=1$, by the formula:$$v_h(q)=dh(q)\lint\hat v(q),\eqno(4.1)$$
where $\lint$ is the interior or contracted product, and $\hat v: \hat Q\to\bigwedge^2 V$ corresponds to $v$. The
formula $\la dh(q),v_h(q)\ra=0$
shows that $v_h$ is tangent to the screen.

4.3. {\sl Restriction of projective trajectories.} The points of ${\cal U}\subset {\cal S}(V)$ are in one-to-one
correspondence with the points of the screen ${\cal U}_h\subset V$ with equation $h=1$. So an oriented path $c:{\cal
I}\to {\cal U}$ restricts into an oriented path $c_h:{\cal I}\to {\cal U}_h$. If $c$ possesses a
structure of projective trajectory, i.e.\ a map $\hat v:{\cal I}\to \bigwedge^2V$ with the tangency condition 3.8,
$c_h$ is endowed with the field of tangent vectors $v_h$; it is a ``usual trajectory''.

4.4. {\sl Law of areas.} If it has no multiple points, a projective trajectory is simply a two-dimensional
semi-cone endowed with a tangent bivector field $\hat v$, which is constant along the rays. The
field
$\hat v$ defines a way to measure the oriented area of any domain delimited on the semi-cone. One takes as unit the
area of a tangent parallelogram spanned by two tangent vectors
$\alpha$ and
$\beta$ such that
$\hat v=\alpha\wedge \beta$.

The usual trajectory $c_h$ on a screen ${\cal U}_h$, restriction of the projective trajectory, is described according
to the ``law of areas'': {\sl the moving point sweeps out from the origin of
$V$ equal areas in equal times}, more precisely it sweeps out a unit of area in two units of time. Indeed,
in a time $dt$ the area swept is half of the area of the parallelogram spanned by $q$ and $v_hdt$. And the area of the
tangent parallelogram spanned by $q$ and $v_h$ is one.

4.5. {\sl The example of the uniform motion.} In 3.10 we stated that a uniform projective trajectory corresponds to a
vectorial plane in
$V$ with a non-zero {\sl constant} area element $\nu\in\bigwedge^2V$, whose support is the plane. From the law of areas
the reader will deduce that the restriction of this projective trajectory to a flat screen is a usual uniform motion,
and its restriction to a screen corresponding to a screen function $h=\|q\|$ is a uniform (geodesic) motion on the unit
sphere
$\|q\|=1$. In this last case we need a positive definite quadratic form $q\mapsto \|q\|^2$. Of course the case of
an indefinite quadratic form is also interesting.

4.6. {\sl A field of forces on a screen.} Given a field of projective forces  on ${\cal U}\subset{\cal S}(V)$ and a
screen function $h:\hat{\cal U}\to\real$, a usual field of forces  may be defined by restriction to
the screen ${\cal U}_h\subset V$ with equation $h=1$. It is tangent to the screen. What kind of dynamics is defined by
such a field of forces?

Let $f_h$ be any vector field tangent to ${\cal U}_h$. {\sl As a force field, it  defines a dynamics through the
system
$$\ddot q=f_h(q)+\lambda q,\eqno(4.2)$$
where the real value of $\lambda$ is forced by the constraint $h(q)=1$}. The term $\lambda q$
is a ``reaction". ``Normal'' reactions are more traditional, but in this framework we
have to take a ``radial'' reaction. The determination  of $\lambda$ is a valuable exercise. As
$h(q)=1$, $\la dh(q),\dot q\ra=0$ and $\la dh(q),\ddot q\ra+\la \partial^2h(q),\dot q\otimes\dot q\ra=0$, where
$\partial^2 h: V\to V^*\otimes V^*$ is the Hessian
quadratic form. By Euler's relation $\la dh(q),q\ra=h(q)=1$. This gives  $\lambda=-
\la \partial^2h(q),\dot q\otimes\dot q\ra$.

It is natural to consider that vector fields are ``simpler" than bivector fields. However, everywhere in this theory
bivectors appear as a simplifying tool. This happens even while working with screen dynamics. Equation $(4.2)$
above becomes $$\dot \pi=f(q),\qquad\hbox{with}\qquad \pi=q\wedge
\dot q,\quad f(q)=q\wedge f_h(q).\eqno(4.3)$$ From this equation we can deduce $(4.2)$: contracting $dh$ at the
left we get $dh\lint(q\wedge\ddot q)=\la dh,q\ra\ddot q-\la dh,\ddot q\ra q=dh\lint f=f_h$. As $\la dh,q\ra=1$, this
equation is of type $(4.2)$.

4.7. {\sl Two screens.} Suppose a field of projective forces is given by a $f:\hat {\cal U}\to\bigwedge^2
V$, positively homogeneous of degree $-2$. Suppose a screen $h(q)=1$ is given, and the force is restricted to the
screen. Suppose a trajectory
$q(t)$ on the screen is found, solution of the screen motion equation
$d\pi/dt=f(q)$, with $\pi=q\wedge (dq/dt)$. Then a projective trajectory is defined extending $\pi(t)$ on the ray of
$q(t)$ with the homogeneity zero.

Suppose
$q_1(t)=\lambda(t)q(t)$, $\lambda(t)\in\real$, is permanently on another screen. If $\tau$ is a new parameter
such that
$d\tau/dt=[\lambda(t)]^{2}$, then $\pi_1(t)=q_1(t)\wedge (dq_1/d\tau)=\pi(t)$. We have
$d\pi_1/d\tau=(d\pi/dt)(dt/d\tau)=\lambda^{-2}f(q)=f(\lambda q)=f(q_1)$. Then $q_1(\tau)$ is a trajectory on the other
screen, solution of the screen dynamics induced by the restriction of $f$ to this screen.

Thus starting with a ray and a projective impulsion, we construct a projective trajectory using a screen; this
trajectory is unchanged if we use another screen.

4.8. {\sl A more intrinsic equation.} Let $t\mapsto \bigl(Q(t),\pi(t)\bigr)$ be a projective trajectory on ${\cal
U}\subset {\cal S}(V)$, with $Q(t)\in {\cal U}$ and $\pi(t)\in\bigwedge^2 V$. There exists a vector
field $v$ along the two-dimensional semi-cone corresponding to $t\mapsto Q(t)$, such that $\pi(t)=q\wedge v$ at any $t$ and
any $q\in \hat Q(t)\subset V$. We should think $v$ as the velocity $\dot q$; it is not unique but determined at $q$ up
to the addition of $\lambda q$, with $\lambda\in\real$. If a field of forces $f$ is given as a field of bivectors on
$\hat{\cal U}$, and if we think $\pi$ as a field of bivectors along the semi-cone, the equation of dynamics may be written:
$$\partial_v\pi=f(q).\eqno(4.4)$$
Here $\partial_v$ means the standard derivation in the direction of the field $v$. Note that this equation coincides
with $(4.3)$, that $\partial_{v+\lambda q}\pi=\partial_v\pi+\lambda\partial_q\pi$ and that $\partial_q\pi=0$ because $\pi$
is homogeneous of degree 0.

4.9. {\sl Remark on a notation.} When we add $v$ to $\lambda q$ we implicitly consider that they are object of the same
type, i.e.\ vector fields along the semi-cone. We will consider $q$ as a notation for the ``Euler field", i.e.\ the
vector field on $V$ whose value at $q$ is $q$. Here this field is implicitly restricted to the semi-cone.
\bigskip

\centerline{\bf 5. Divergence-free fields of projective forces}

To study the divergence operator acting on vector fields,  we first need to introduce the generalized differential
forms on
${\cal U}\subset {\cal S}(V)$. We build up an algebra of tensors on the $s$-tangent vectors introduced in \S 2. We do
not fix a terminology and try to work with a minimal system of notation. We adopt a simplified notation for
$T^0{\cal U}$ and forget
the notation $T^s{\cal U}$, using instead:
$${\cal X}{\cal U}=T^{0}{\cal U},\qquad {\cal X}{\cal U}\otimes \theta^s{\cal U}=T^s{\cal U}.$$
The generalized tensors fields are the sections of $\bigotimes^i{\cal X}^*{\cal U}\otimes\bigotimes^j{\cal X}{\cal
U}\otimes\theta^s{\cal U}$. The ``usual'' tensors fields, i.e.\ those constructed considering that ${\cal U}$ has only
the local structure of a differentiable manifold, are sections of $\bigotimes^i T^*{\cal
U}\otimes\bigotimes^jT{\cal U}$. But $T{\cal U}=T^1{\cal U}={\cal X}{\cal U}\otimes\theta^1{\cal U}$ and $T^*{\cal U}
={\cal X}^*{\cal U}\otimes \theta^{-1}{\cal U}$; expanding this we get $\bigotimes^i T^*{\cal
U}\otimes\bigotimes^jT{\cal U}=\bigotimes^i{\cal X}^*{\cal U}\otimes\bigotimes^j{\cal X}{\cal
U}\otimes\theta^{j-i}{\cal U}$.

A $\omega\in \Gamma(\bigwedge^i{\cal X}^*{\cal U}\otimes \theta^s{\cal U})$ {\sl corresponds} to a
$\hat\omega:\hat{\cal U}\to
\bigwedge^i V^*$, positively homogeneous of degree $s$, and such that $q\lint \hat\omega(q)=0$ for any $q\in\hat U$.
The following identities hold at the point
$q\in\hat{\cal U}$:
$$q\lint d\hat\omega={\cal L}_q\hat\omega=(i+s)\hat\omega.\eqno(5.1)$$
The first is Cartan formula. Let $\partial_q$ be the standard derivation, in the
direction of the Euler field $q$, of functions defined on $V$. We get, using coordinates or considerations on
the derivations,
${\cal L}_q\hat\omega-\partial_q\hat\omega=i\hat\omega$. But by Euler's characterization of homogeneity
$\partial_q\hat\omega=s\hat\omega$, which gives the second formula.

5.1. {\sl Proposition.} Let  $\omega\in \Gamma(\bigwedge^i{\cal X}^*{\cal U}\otimes \theta^s{\cal U})$ and
let $\hat\omega:\hat{\cal U}\to
\bigwedge^i V^*$ be the corresponding $i$-form. Suppose there exist $s'\in\real$ and  $\rho\in
\Gamma(\bigwedge^{i+1}{\cal X}^*{\cal U}\otimes
\theta^{s'}{\cal U})$ such that $d\hat\omega=\hat\rho$. Then $i+s=0$: $\omega$
is a ``usual''
$i$-form on
${\cal U}$ and consequently possesses an exterior derivative $d\omega$. Finally $d\omega=\rho$ and $s'=s-1=-i-1$.

{\sl Proof.} Assuming $d\hat\omega=\hat\rho$, we get $q\lint d\hat\omega=0$ and $i+s=0$ by $(5.1)$.
Finally $d\omega$ satisfies $d\hat\omega=\hat{d\omega}$, because $\hat{\  }$ is
the pull-back operation by the canonical projection $\hat{\cal U}\to{\cal U}$, which commutes with $d$.

{\sl Divergence-free fields of forces.} Let $h:\hat {\cal U}\to \real$ be a screen function, and $f_h:{\cal U}_h\to V$
be a vector field tangent to the screen ${\cal U}_h$ with equation $h=1$. If we choose a unit of volume on $V$,
an area form is canonically defined on ${\cal U}_h$ and the divergence of $f_h$ is well-defined.

Let $\mu\in\bigwedge^{n+1}V^*$ be the non-zero volume form defining the unit of volume in $V$. The $n$-dimensional area
form at
$q\in {\cal U}_h$ is simply the restriction to ${\cal U}_h$ of the form $q\lint\mu$. One can also think of it more
geometrically. The area of an open set in ${\cal U}_h$ is $n+1$ times the volume of the cone with base the open set and
vertex at the origin of $V$.

The divergence of $f_h$ is the function on ${\cal U}_h$ defined by the classical identity $({\rm div}f_h) q\lint
\mu=d(f_h\lint(q\lint \mu))$. We set $f=q\wedge f_h$; then $f_h\lint(q\lint\mu)=-f\lint\mu$. We extend $f$
by positive homogeneity of degree
$-2$ in a projective force field $f:\hat {\cal U}\to\bigwedge^2 V$. {\sl The vector field
$f_h$ is divergence-free if and only if the restriction to ${\cal U}_h$ of the
$n$-form $d(f\lint\mu)$ vanishes identically.}

This last condition is not invariant under a change of screen. However, if $d(f\lint\mu)$ vanishes
identically on a screen ${\cal U}_h$, and if moreover $q\lint d(f\lint\mu)=0$, then it vanishes identically on $\hat{\cal
U}$. {\sl It is only in this case that we can say that the projective force field $f$ is divergence-free,} i.e.\
${\rm div}f_h=0$ for {\sl any} screen function
$h$. We can
apply Proposition 5.1 to the closed form
$\hat\omega=f\lint\mu$, which corresponds to a section $\omega$ of $\bigwedge^{n-1}{\cal X}^*{\cal
U}\otimes\theta^{-2}{\cal U}$. This gives the following statement.

5.2. {\sl Proposition.} Divergence-free projective force fields exist only if $n=3$, i.e.\ if $\dim V=4$. In this
dimension, if we choose a unit of volume on $V$, the space of projective force fields on ${\cal U}\subset{\cal S}(V)$
is canonically identified with the space of
$2$-forms on ${\cal U}$. A divergence-free projective force field is sent on a  closed $2$-form. In this dimension
again, if a projective force field restricts to a given screen ${\cal U}_h$ into a divergence-free vector field,
it is a divergence-free projective force field, and consequently it restricts to any screen into a divergence-free
vector field.

\bigskip
\centerline{\bf 6. Generalized Kepler problem}

6.1. {\sl Central forces.} Let $V$ be a vector space of dimension $n+1$, $h\in V^*$ a non-zero linear form, and
$A\subset V$ the affine hyperplane with equation $h=1$. Let $c\in A$ be the ``center of force''. We define the motion of
$q\in A\setminus{\{c\}}$ in a central force field by the system
$$\ddot q=-\psi(q)(q-c),\qquad \psi: A\setminus{\{c\}}\to\real.\eqno(6.1)$$
Let ${\cal U}\subset {\cal S}(V)$ be the set of rays intersecting $A\setminus{\{c\}}$, and $\hat{\cal U}$ the
corresponding semi-cone. We extend by positive homogeneity the force field to obtain a projective force field
$f:\hat{\cal U} \to\bigwedge^2V$, whose expression is $$f(q)=\Psi(q)q\wedge c,\qquad\hbox{where}\qquad \Psi(q)=\la
h,q\ra^{-3}\psi\Bigl(\frac{q}{\la h,q\ra}\Bigr)\eqno(6.2)$$ is positively homogeneous of degree $-3$ and satisfies
$\Psi(q)=\psi(q)$ on $A$.
 
6.2. Let us suppose that $\psi$ has the homogeneity property of the coefficient $\|q-c\|^{-3}$
in the Kepler problem; namely
$$\psi(q)=\phi(q-c),\quad\hbox{where}\quad \phi(\lambda v)=\lambda^{-3}\phi(v),\quad\lambda>0,\quad \la
h,v\ra=0.\eqno(6.3)$$ Then $\Psi(q)=\phi(q-\la h,q\ra c)=\phi\bigl(h\lint (c\wedge q)\bigr)$. We have $\Psi(q+\gamma
c)=\Psi(q)$ for any
$(q,\gamma)\in\hat {\cal U}\times\real$  such that $q+\gamma c\in\hat {\cal U}$. {\sl The
coefficient
$\Psi$ is invariant by any translation with direction $c$. Let  $[c]\subset V$ be the line generated by $c$. The
coefficient $\Psi$ and consequently the projective force field are extended canonically to $V\setminus [c]$ by this
property.}

This canonical extension exists for the Kepler problem (see \S 7; observe that $\Psi$ is an even
function of $q$.) One can also extend force fields defined by several fixed point masses on
$A$, or more generally by any fixed repartition of mass. We will only study the situation above, i.e.\ the
motion in a projective force field
$$f(q)=\Psi(q)q\wedge c,\eqno(6.4)$$
where for any $q\in V\setminus[c]$, $\gamma\in\real$ and $\lambda>0$:
$$\Psi:V\setminus[c]\to\real,\quad\Psi(q+\gamma
c)=\Psi(q),\quad\Psi(\lambda q)=\lambda^{-3}\Psi(q).$$

6.3. {\sl Lemma.} Let $G$ be the group of linear transformations of
$V$ that fix
the vector $c$ and induce the identity on the quotient space
$V/[c]$. An element of $G$ sends any projective trajectory solution of Problem $(6.4)$ onto another solution.

{\sl Proof.} The projective force field $f$ of $(6.4)$ is trivially invariant by such a transformation.

6.4. {\sl Lemma.} The bivector ``constant of areas'' $C=(q-c)\wedge \dot q$ is invariant along the solutions of
$(6.1)$. It is ``projectivized'' in ${\cal C}=c\wedge \pi\in \bigwedge^3 V$, where $\pi=q\wedge\dot q\in\bigwedge^2 V$ is
the projective impulsion: ${\cal C}$ is constant along the projective trajectories defined by the projective force field
$(6.2)$. On the other hand, the linear transformations of the group $G$ introduced in Lemma 6.3 preserve ${\cal C}$.

{\sl Proof.} Using the equation of dynamics $(4.4)$, we see that ${\cal C}$ is invariant along a projective trajectory:
$\partial_{\dot q}{\cal C}=c\wedge\partial_{\dot q}\pi=c\wedge f=\Psi(q)c\wedge q\wedge c=0$. Let $g\in G$. We decompose
$\pi$ arbitrarily:
$\pi=q\wedge q'$. As $g(q)=q+\gamma c$ and $g(q')=q'+\gamma'c$ for some
$(\gamma,\gamma')\in\real^2$, we have ${\cal C}=c\wedge q\wedge q'=c\wedge g(q)\wedge g(q')$.

{\sl Two-dimensional dynamics.} The dynamics of $(6.1)$ and of the projectivized versions $(6.2)$ and $(6.4)$
are essentially 2-dimensional. We will content ourselves with a study of the case $n+1=\dim V=3$. Hypothesis
$(6.3)$ corresponds to what is called in [AlS] a Jacobi attractor. Some conditions are given there ensuring that under this
hypothesis,
$(6.1)$ possesses open sets filled by periodic orbits. Let us show how we arrive at the
conclusions of [AlS] about the dynamics around a Jacobi attractor by an elementary study of the action of the group
$G$.

6.5. {\sl Lemma.} The group $G$ of Lemma 6.3 is abelian and canonically isomorphic to the additive group $[c]^0\subset
V^*$, the annulator of the ``vertical" line
$[c]$. Assume
$\dim V=3$. The domain $\hat {\cal U}=V\setminus [c]$ is foliated by vertical half-planes
with boundary the vertical line $[c]$ that we simply call the {\sl leaves}. Let us fix such a leaf $\hat L\subset V$
and call $L\subset {\cal S}(V)$ the corresponding object; $L$ is invariant by the action of $G$. We call $L_{\cal
C}\subset T^{-1}{\cal U}$ the set of projective states with position in $L$ and given constant of areas ${\cal C}\neq
0$; then $G$ acts simply transitively on $L_{\cal C}$, giving to $L_{\cal C}$ the structure of an affine plane, and
to the submanifold of states in
$T^{-1}{\cal U}$ with constant of areas ${\cal C}$ the structure of a principal $G$-bundle with base the
one-dimensional manifold
${\cal S}(V/[c])$.

{\sl Proof.} The elements of $G$ have the form $\1+\omega\otimes c\in V^*\otimes V$, where $\omega\in V^*$ satisfies
$\la\omega,c\ra=0$. The first statement follows easily. Let us fix a leaf $\hat L$, and choose linear coordinates
$(x,y,z)$ of $V$ such that $c=(0,0,1)$ and $\hat L$ is $y=0$, $x>0$. To give an element of $L_{\cal C}$, we give a $q$
with $y(q)=0$ that we normalize with the condition $x(q)=1$, then we give a $\dot q$ such that $q\wedge\dot q$ is the
projective impulsion. We choose $\dot q$ with $x(\dot q)=0$, and we must take $y(\dot q)=\la dx\wedge dy\wedge
dz,{\cal C}\ra$. After these normalizations only the $z$-components of $q$ and $\dot q$ are free. But 
$g_{\gamma\gamma'}=\1+(\gamma dx+\gamma'dy)\otimes c\in G$ sends $(q,\dot q)$ onto $(q+\gamma c,\dot q+\gamma'
c)$.

6.6. {\sl Proposition.} Consider Problem $(6.4)$, defined on the domain $\hat {\cal
U}=V\setminus [c]$. Assume moreover
$\dim V=3$, fix an integer $k\geq 0$ and a constant of areas ${\cal C}\neq 0$. Consider a ``source'' leaf $\hat
L\subset V$ and  ``target'' leaf $\hat K\subset V$. Any initial condition in $L_{\cal C}$ (defined in 6.5) defines a
projective trajectory that cuts the leaf $K$ infinitely many times ``afterwards''. The map from $L_{\cal C}$ to
$K_{\cal C}$ that associate to the initial condition in $L_{\cal C}$
the state at the intersection with $K$ after $k$ turns commutes with the action of $G$.

{\sl Proof.} The domain ${\cal U}\subset {\cal S}(V)$ is topologically the sphere minus two points, the ``poles'', and a
trajectory with
${\cal C}\neq 0$ cuts the projected leaves, the ``meridians'', transversally. We prove first that any such trajectory is
made of an infinity of ``loops'' around ${\cal U}$. If it was not the case, the trajectory would stop at a pole, and by
monotonicity of the ``longitude'', would do it tangentially to a meridian. But in a small sector near this
meridian, the field of forces is close to the field of forces in a Kepler problem, for which such behavior is
impossible. So the trajectory must turn indefinitely in the past and in the future. The commutation statement is Lemma
6.3.

6.7. {\sl Eccentricity vector.} For any orbit with ${\cal C}\neq 0$, there is a unique element $g\in G$
that sends a state in  $L_{\cal C}$ to the ``next'' intersection of the orbit with $L$. If $g\neq \1$ then the space
of orbits with same ${\cal C}$ is a cylinder, isomorphic to the quotient of $G$ by the subgroup generated by $g$, and
these orbits are not closed. If $g=\1$ we are in what we called in [AlS] and [Al1] the Jacobi-Darboux case. The
orbits are closed. In term of the
$G$-principal bundle described in 6.5, Problem $(6.4)$ defines a local trivialization of the bundle, and in
the case $g=\1$ this local trivialization has no monodromy.

This means that we got a first integral for the states with given ${\cal C}$. When $g=\1$ it takes value in one of the
fibers
$L_{\cal C}$. If we choose a reference orbit with this angular momentum\footnote{In the Kepler problem, for which $g=\1$,
the natural choice is the circular orbit.}, the value of the first integral may be identified to the element of $G$ that
sends the reference orbit onto the present orbit. The element of $G$ are covectors (see 6.5). The covectorial value of the
first integral is related to the well-known eccentricity vector.

We have all in mind that to a first integral should be associated a symmetry. Nothing excludes that in the same problem
different ``associations" coexist. Here we can speculate that to the action of the abelian symmetry group
$G$ is associated a ``moment'' that may be an element of the Lie algebra of $G$, and may be the
eccentricity vector.

If we pass to the particular case of the Kepler problem, which has a natural Hamiltonian structure, this association
coexists with the Hamiltonian association, where the ``moment" is the eccentricity vector multiplied by the square root of
the given semimajor axis, and the Hamiltonian action is the rotation of the Bacry-Gyorgyi parameters (as defined in [Sou]
or [Cor]).

Up to now we gave to ${\cal C}$ a fixed non-zero value. Is the dynamics similar for the other non-zero values of ${\cal
C}$? The Jacobi problem
$(6.3)$ in affine dynamics possesses an obvious invariance by dilation $(q-c,\dot q)\mapsto (\lambda
(q-c),\lambda^{-1/2}\dot q)$, which does not preserve the time parameterization. From the projective view point, this
transformation is defined after the choice of an affine screen, and it does not commute with $G$. Nevertheless it extends
to ${\cal U}$ and preserves the leaves. If we call
$(z,\dot y,\dot z)=(z(q),y(\dot q),z(\dot q))$ the coordinates after the normalization in the proof of 6.5, the map is
$(z,\dot y,\dot z)\mapsto (\lambda^{-1}z,\lambda^{-1/2}\dot y,\lambda^{-3/2}\dot z)$. This map may be used to establish
that if the orbits are closed for a given non-zero value of ${\cal C}$, they are closed for any non-zero value of ${\cal
C}$.

\bigskip
\centerline{\bf 7. The Kepler problem}

If $(6.1)$ is $(1.4)$ with $\beta=-3$, i.e.\ $\ddot q=-\|q-c\|^{-3}(q-c)$, the properties obtained in the previous
section apply. The field of projective forces extends to $V\setminus [c]$. Its expression is:
$$f(q)=\|q\|^{-3}q\wedge c,\eqno(7.1)$$
where $\|.\|^2$ is a positive quadratic form on $V$, satisfying $\|c\|^2=0$ and having only $[c]$ as a direction of
degeneracy.

7.1. {\sl Proposition.} Let $H$ be the group of linear transformations of $V$ that preserve the degenerate
quadratic form
$\|.\|^2$ and fix
$c$. Let $\dim
V=n+1$. The dimension of $H$ is $n(n+1)/2$. An
element of $H$ sends any projective trajectory solution of $(7.1)$ onto another solution, and the constant of
areas ${\cal C}$ is preserved.

{\sl Proof.} Let us choose a vectorial hyperplane $W\subset V$ that does not contain $c$, and $l\in H$. The image $l(W)$
of $W$ is a hyperplane that does not contain $c$. But the group $G$ of Lemma 6.3 is a subgroup of $H$. We know
that it possesses a unique element $g$ that sends $W$ on $l(W)$. Now $g^{-1}\circ l=k$ is
an isometry of $W$ for the induced Euclidean form. Conversely any $g\in G$ composed with any isometry $k$ of $W$ is
an element of
$H$. Then
$\dim H=\dim O(W)+\dim G=n(n-1)/2+n$. The last statements come from the invariance of $f$ under the action of $H$, and
from the invariance of ${\cal C}$ stated in 6.4.

Problem $(7.1)$ may be seen as the ``abstract'' Kepler problem. The usual Kepler problem in the Euclidean space, and its
19th century generalizations to spaces of constant curvatures, appear as ``materializations'' obtained by
choosing different screens. We will come back to these
screens in a forthcoming work\footnote{See [BoM] for these generalizations. The remark that these different Kepler problems
are related by central projection and change of time is due to Appell (see [Ap1], p.\ 158).}. They all remove part of the
symmetry of the abstract problem.

But there is also a canonical screen: the screen $\|q\|=1$, the ``unit cylinder'' of the degenerate quadratic form.
The group $H$ is a symmetry group for this cylinder. In a sense $\|q\|=1$ is more symmetric than a usual Euclidean
cylinder, because it admits the subgroup
$G$ of 6.3, whose elements are called ``transvections'', as a group of symmetry. But in another sense it is less
symmetric, because the translations in the
$c$ direction are not in the symmetry group. Restricting Problem $(7.1)$ to this screen we ``materialize'' the abstract
Kepler problem without removing any symmetry.

\espacefig
\centerline{\includegraphics [width=100mm] {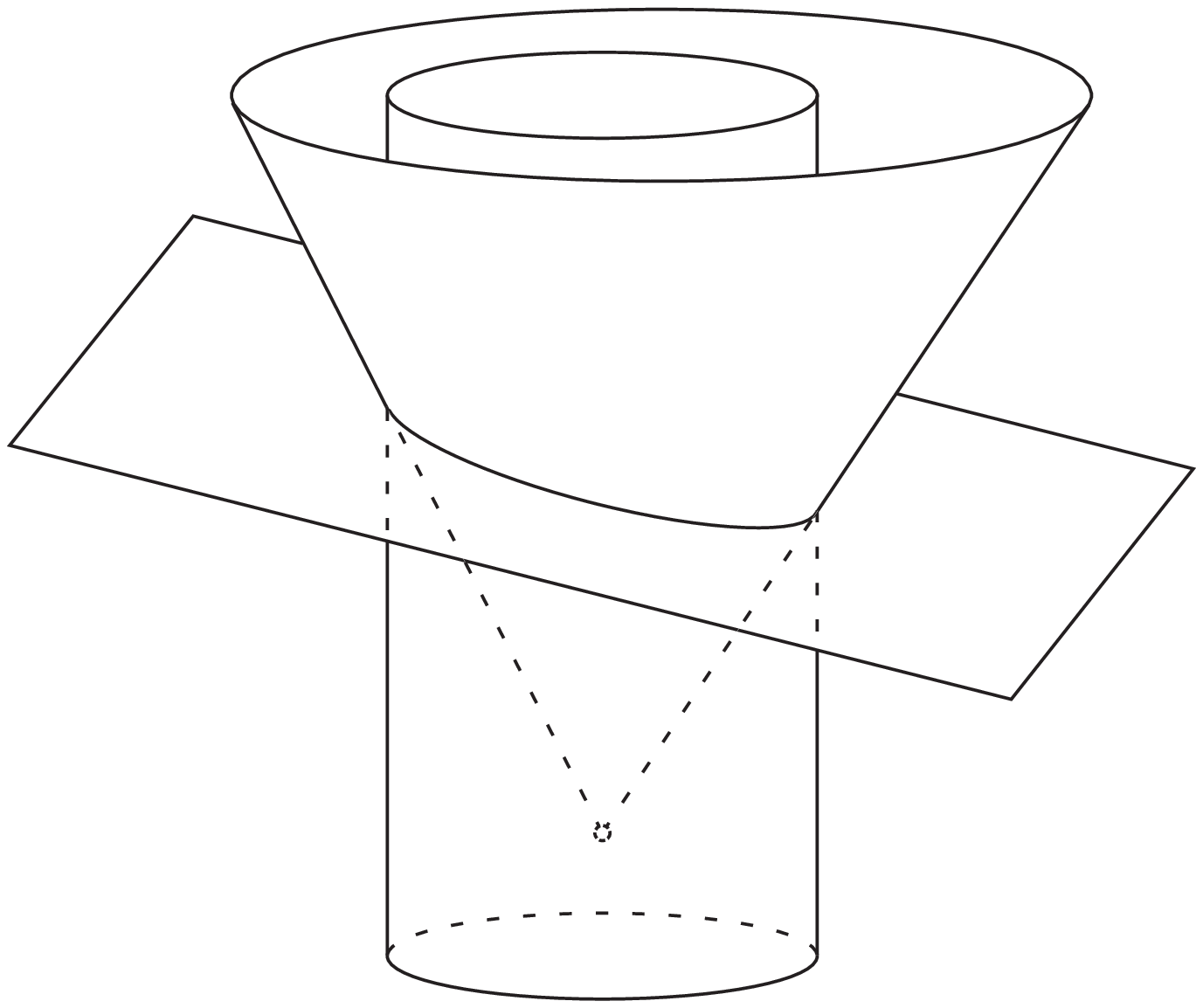}}
\nobreak
7.2. {\sl Figure.} Construction of a Keplerian cone.

As usual we think of $c$ as the ``vertical'' direction. The field of forces induced on the screen by restriction of
$(7.1)$ is vertical and constant: the screen function is $h(q)=\|q\|$, so on $h(q)=1=\la dh,q\ra$ the restricted force
$f_h=dh\lint f=\la dh,q\ra  c-\la dh,c\ra q$ is $c$. The next proposition will describe the orbits.
But we begin with the free motion, i.e.\ the case of a zero force.

We know that a free motion describes the intersection of the cylinder $\|q\|=1$ with a vectorial plane. This does not
look like a free motion on a usual cylinder, but one should not forget that in screen dynamics, the reaction, that
maintains the particle on the screen, is ``radial'' and not normal (see \S 4.6). The screen dynamics equation for the
free motion is $\ddot q=\lambda q$, where $\lambda$ is imposed by the constraint $\|q\|=1$. If we project vertically
this equation, i.e.\ if we project by the canonical projection $V\to V/[c]$, $q\mapsto q_c$,
the equation remains of the same form $\ddot q_c=\lambda q_c$, which implies that the ``horizontal motion'' is uniform
(the reaction being normal to the circle $\|q\|=\|q_c\|=1$). Let us call $\dot \theta$ the constant angular velocity of
this motion. Then $\lambda=-\dot\theta^2$.

7.3. {\sl Proposition.} The screen dynamics of the Kepler problem $(7.1)$ restricted to the cylindrical screen $\|q\|=1$
is as follows. If  ${\cal C}=0$ the particle describes the vertical lines (with
direction $c$) in a uniformly accelerated motion. If ${\cal C}\neq 0$ the particle moves along the
intersection of the cylinder with a non-vertical affine plane, and does it in such a way that the canonical angular
variable on the cylinder increases or decreases uniformly.

{\sl Proof.} The equation is $\ddot q=\lambda q+c$. The constant of areas is ${\cal C}=c\wedge q\wedge\dot q$. As in the
case of a free motion, $\dot \theta$ is a constant proportional to ${\cal C}$ and $\lambda=-\dot\theta^2$. If we
set $q=\dot\theta^{-2}c+q_0$, the equation becomes $\ddot q_0=\lambda q_0$, equation of the free motion. The
conclusion follows from the study of the free motion. The affine plane passes through the
vector
$\dot\theta^{-2}c$.

The intersection of the affine plane with the cylinder is a conic section. {\sl In a projective trajectory of the
Kepler problem the ray-particle describes a quadratic semi-cone}. The most famous result of Newton's {\sl Principia},
that the orbits of the usual Kepler problem are conic sections, is an obvious corollary. The above
deduction of this famous result is indeed a kind of geometrization of the well-known deduction where one
uses as variables the inverse of the distance from the particle to the center and the polar angle $\theta$.

7.4. {\sl About the parameterization of the paths.} A projective trajectory possesses a field of tangent projective
impulsions. Does it possess also, canonically, a field of tangent velocity vectors? In another way (see 3.3), is it
canonically a usual trajectory? The answer is no for the uniform motion, but yes for the Keplerian motion: the cylindrical
screen is canonical. It converts canonically the field of projective impulsions into a field of velocity vectors.
But still the general
``chronological" structure is the tangent field of projective impulsions, even if in particular
situations some classes of parameterizations may appear as concomitant to the given field of projective forces.

Let us visualize the ``web" of all the projective lines drawn in the projective space.
A point and a tangent direction uniquely determine a line. The ``projective geometry of paths", a classical subject
founded by the papers of Weyl in 1921 and Eisenhart in 1922, describes some possible ``deformations" of this web, where
a point and a tangent direction still determine a curve. It is possible to introduce in this theory the concept
of projective impulsion. A projective impulsion is more than a tangent direction. However changing
the length of the projective impulsion without changing its direction we do not change the curve.

An extension to the projective geometry of paths has been investigated by Tabachnikov (see [OvT]): the ``projective
billiards". Here also the notion of projective impulsion could be introduced.
We could add to the model a field of projective forces, thus ``deforming" again the web of curves. Then the curves with
same initial direction but distinct initial projective impulsion separate into distinct trajectories.

There is a deep relation between the existence of interesting reparameterizations and the integrability of a system.
The discoveries by Matveev, Topalov [MaT] and Tabachnikov [Tab] on the geodesic flow on the ellipsoid in any dimension,
and their subsequent investigations on the subject are remarkable (see [Ta1]).

To come back to our initial question, let us mention that several works on the
projective geometry of paths discuss parameterizations
or classes of parametrizations of the paths. Works by T.Y.\ Thomas and by Whitehead are often mentioned. However, as
far as we know, these constructions are not canonical, and involve more structure than what is strictly necessary to
define the paths.

In forthcoming papers we wish to insist on the notion of projective impulsion. We can do this by
presenting the focal theory of conic sections in pseudo-Riemannian spaces of constant curvature, in such a way that
their role in the Kepler problem is deduced very simply. We will also explain some results on the first
integrals of the uniform motion, which are useful in the study of projective dynamics. Our projective
impulsion point of view clarifies a classical literature on the subject culminating with a result of
Nijenhuis\footnote{[Nij]. For shorter proof see [Th1]. Many thanks to Bartolom\'e Coll who introduced me to this
literature.}.

\bigskip
\centerline{\bf 8. Bibliographical Notice}

{\sl Halphen's remark.} In 1878, Georges-Henri Halphen explained some of the ideas we presented in 1.1 and 1.3,
beginning with these words: ``On me permettra, en dernier lieu, d'appeler l'attention sur ce fait curieux..." This
is, as far as we know, the first time a projective transformation is considered in dynamics.

Halphen introduced a projective transformation in order to synthesize his answer to a question raised by Bertrand.
Bertrand wanted to know what are the fields of forces in an affine space, depending only of the position, such
that the particle describes a conic section, whatever be the initial condition.

In 1877, a week after the publication of his question, Bertrand announced that he had just completed a first step toward
the solution: he proved that the force had to be central. Darboux attended his lecture at the Coll\`ege de
France and immediately completed the solution. Soon after Halphen presented his independent resolution\footnote{The
references are the volume 84 (1877) of Comptes rendus de l'Acad\'emie des Sciences, pp.\ 673, 731, 760, 936, 939, and
[Hal] for the remark on projective transformations. Actually, it is not clear that Bertrand's problem has ever been
completely solved. An objection appears in a parenthesis of [Rou], \S 393 ``unless the force at every point of that curve
is infinite". The context of the original question makes clear that Bertrand looks for all the force fields in the plane.
To deduce that the force field is central, we must use Bertrand's argument or Halphen's argument. Halphen's argument is
3-dimensional and does not work in the plane. Bertrand's argument does work and shows that the lines of force are straight
lines. But these lines could be, for example, the tangent lines to a strictly convex domain. The force field would only be
defined out of this domain. In [Imc] there is the claim of a complete solution of Bertrand's question, but the argument is
clearly wrong. If we start at p.\ 32 with a force field $(X,Y)$ which is central and satisfies Bertrand's requirement,
but whose center is not the origin of the coordinates, we conclude at p.\ 36 that the origin is also a center of force,
which is absurd.}.

Darboux and Halphen found two classes of solutions, and Halphen found a wonderful description of these classes: they are
the field of Newtonian attraction from a center {\sl and all its affine transforms} and the field of linear forces
(i.e.\ $\ddot q=-q$) from a center {\sl and all its projective transforms}. In fact we should accept imaginary
parameters in the transformations to allow changes in the signature, or keep everything real and complicate a bit the
statement. But our simplified statement raises immediately the question: why not the projective transforms of the
Newtonian attraction also? And the similar but easier question about the linear force? The answer to the first question is:
the Newtonian attraction is invariant by the projective transformations. This is the starting point of projective
dynamics.

Some earlier works by Newton
and Hamilton describing the second class were mentioned by Glaisher (see also [Tis] p.\ 42). But apparently they were
not followed by a remark on the projective transformation.

The articles [App] and [Ap1] by Paul Appell on the central projection
present several results, including some characterization of the projective transformations within a larger class of
transformations acting on the space-time. Kasner later improved this characterization and extended it with De Cicco.

Appell's work was mainly continued through a question suggested by Goursat (see the end of [App]) and the papers of
Painlev\'e, Ren\'e Liouville and Levi-Civita dedicated to it (see also [Lut]). These works remained quite far
from the original remark by Halphen. Also in my sense the problem stated by Goursat does not respect the ``non-quadratic''
or affine character of equation $(1.1)$. As far as I know, all the related and subsequent works involve a metric, even if
T.Y.\ Thomas happily joined the question to the projective geometry of paths, and stated it using a connection (see
[Tho] and [LiA]).

{\sl Acknowledgements.} We wish to thank Alain Chenciner, Mauricio Garay
and Richard Montgomery for their very stimulating comments.

\bigskip
\centerline{\bf 9. References}

[Alb] A. Albouy, {\sl Lectures on the two-body problem}, in ``Classical and Celestial Mechanics. The Recife
Lectures.'' edited by H.\ Cabral and F.\ Diacu,  Princeton University Press (2002)
pp.\ 63--116

[Al1] A. Albouy, {\sl The underlying geometry of the fixed centers problems}, in ``Topological
Methods, Variational Methods and their applications'', edited by H.\ Brezis,
K.C.\ Chang, S.J.\ Li and P.\ Rabinowitz, World Scientific (2003) pp.\ 11--21

[AlC] A. Albouy, A. Chenciner, {\sl Le probl\`eme des N corps et les
distances mu\-tu\-el\-les}, Inventiones Mathematicae 131 (1998)
pp.\ 151--184

[AlS] A. Albouy, T. Stuchi, {\sl Generalizing the classical fixed-centres problem in a non-Hamil\-to\-nian way}, J.\
Phys.\ A: Math.\ Gen.\  37 (2004) pp.\ 9109--9123

[App] P. Appell, {\sl De l'homographie en m\'ecanique},
American Journal of Mathematics 12 (1890) pp.\ 103--114

[Ap1] P. Appell, {\sl Sur les lois de forces centrales faisant 
d\'ecrire \`a leur
point d'applica\-tion une conique quelles que soient les conditions initiales},
American Journal of Mathematics 13 (1891) pp.\ 153--158

[BoM] A.V. Borisov, I.S. Mamaev, {\sl Generalized problem of two and four
Newtonian centers}, to appear in Celestial Mechanics \& Dynamical Astronomy;
{\sl Superintegrable systems on a sphere}, preprint;
{\sl Classical dynamics in non-Euclidean spaces},
Ser.\ Modern Cel.\ Mech., Moscow-Izhevsk, ICS.\ 2004 (in Russian)

[Cor] B. Cordani, {\sl The Kepler Problem
Group Theoretical Aspects, Regularization and Quantization, with Application to the Study of Perturbations},
Birk\-h\"au\-ser (2003)

[Eis] L.P. Eisenhart, {\sl Spaces with corresponding paths}, Proc.\ Nat.\ Acad.\ 8 (1922) pp.\ 233--238

[Eul] J.A. Euler, {\sl Sur le tems de la chute d'un corps attir\'e vers un centre fixe en raison r\'eciproque des
distances},
 (1760) L.\ Euleri opera omnia II-6, Lausannae (1957) pp.\ 294--302

[Gla] J. Glaisher, {\sl On the Law of Force to any Point in the Plane of Motion, in order that the Orbit may be always a
Conic}, Monthly Notices of the Royal Astronomical Society, 39 (1878) pp.\ 77--91

[Hal] G.H. Halphen, {\sl Sur les lois de Kepler}, Bulletin de la Soci\'et\'e Philomatique de Paris 7-1 (1878) p.\ 89,
\oe uvres 2, Gauthier-Villars (1918) p.\ 93

[Imc] V.-G. Imchenetsky, {\sl D\'etermination en fonction des coordonn\'ees de la force qui fait mouvoir un point mat\'eriel
sur une section conique,} M\'emoires de la soci\'et\'e des sciences de Bordeaux 2-4 (1880) pp.\ 31--40, traduction d'un
article paru \`a Kharkof (1879)

[Kas] E. Kasner, {\sl Differential-geometric aspects of dynamics}, Amer.\ Math.\ Soc.\ Colloquium Publications, vol.\ 3
(1913, 1934), reprint in: G.C.\ Evans, {\sl The logarithmic potential and other monographs}, Chelsea (1980)

[KDC] E. Kasner, J. De Cicco, {\sl Generalization of Appell's transformation}, Journal of mathematics and physics 27
(1949) pp.\ 262--269

[Lev] T.\ Levi-Civita, {\sl Sulle trasformazioni delle equazioni dinamiche}, Annali di Matematica 2-24 (1896) pp.\ 255-300

[LiA] A. Lichnerowicz, D. Aufenkamp, {\sl The general problem of the transformation of the equations of dynamics},
J.\ Ration.\ Mech.\ Anal.\ 1 (1952) pp.\ 499--520

[Lio] R. Liouville, {\sl Sur les \'equations de la dynamique}, Acta Mathematica 19 (1895) pp.\ 251--283

[Lut] J. L\"utzen, {\sl Interactions between Mechanics and Differential Geometry in the 19$^{th}$ Century}, Archive for
History of Exact Sciences 49 (1995) pp.\ 1--72

[MaT] V.S. Matveev, P.J. Topalov, {\sl Trajectory equivalence and corresponding integrals}, Regular and Chaotic Dynamics 3
(1998) pp.\ 30--45

[Mon] R.\ Montgomery, {\sl Hyperbolic Pants fit a three-body problem}, preprint

[New] I. Newton, {\sl Mathematical Principles of Natural Philosophy},
(1687--1713--1726) Motte's translation revised by F.\ Cajori, University of
California Press, Berkeley (1934), or A New translation by I.B.\ Cohen and A.\
Whitman, University of California Press (1999) book 3, general scholium

[Nij] A. Nijenhuis, {\sl A note on first integrals of geodesics}, Proc.\ Kon.\ Ned.\ Akad.\ v.\ Wetens.\ Ser.\ A, 52
(1967) pp.\ 141--145

[OvT] V. Ovsienko, S. Tabachnikov, {\sl Projective differential geometry, old and new: from Schwarzian derivative to
cohomology of diffeomorphism groups}, Cambridge Univ.\ Press (2004)

[Pai] P. Painlev\'e, {\sl M\'emoire sur la transformation des \'equations de la Dynamique}, Journal de math\'ematiques pures
et appliqu\'ees, 10 (1894) pp.\ 5--92

[Rou] E.J. Routh, {\sl A treatise on dynamics of a particle}, Cambridge University Press (1898), Dover (1960)

[Sou] J.-M. Souriau, {\sl Sur la vari\'et\'e de Kepler}, symposia mathematica 14 (1974) pp.\ 343--360

[Tab] S. Tabachnikov, {\sl Projectively equivalent metrics, exact transverse line fields and the geodesic flow on the
ellipsoid}, Comment.\ Math.\ Helv.\ 74 (1999) pp.\ 306-321

[Ta1] S. Tabachnikov, {\sl Ellipsoids, complete integrability and hyperbolic geometry}, Moscow Math.\ J.\ 2 (2002) pp.\
185--198

[Tis] F.-F. Tisserand, {\sl Trait\'e de m\'ecanique c\'eleste}, Tome 1 Gauthier-Villars, Paris (1889), r\'eimpression Gabay
(1990)

[Tho] T.Y. Thomas, {\sl On the transformation of the equations of dynamics}, Journal of mathematics and physics 25 (1946)
pp.\ 191--208

[Th1] G. Thompson, {\sl Killing tensors in spaces of constant curvature}, J.\ Math.\ Phys.\ 27 (1986) pp.\ 2693--2699

[Wey] H. Weyl, {\sl Zur Infinitesimalgeometrie: Einordnung der projektiven und der konformen Auffassung}, G\"ottinger
Nachrichten (1921) pp.\ 99--112

\end{document}